\documentclass{emulateapj}
\usepackage{lineno}
\usepackage{natbib}
\usepackage{hyperref}
\usepackage{rotating}
\usepackage{multirow}
\usepackage{subfigure}
\shorttitle{VERITAS Observations of Gamma-Ray Bursts Detected by {\it Swift}}
\shortauthors{Acciari et al.}

\newcommand{\degree}{\ensuremath{^\circ}} 
\newcommand{\hour}{\ensuremath{^h}} 
\newcommand{\minute}{\ensuremath{^m}} 
\newcommand{\second}{\ensuremath{^s}} 

\begin{document}

\title{VERITAS Observations of Gamma-Ray Bursts Detected by {\it Swift}}

\author{ V.~A.~Acciari\altaffilmark{1}, E.~Aliu\altaffilmark{2}, T.~Arlen\altaffilmark{3},
  T.~Aune\altaffilmark{4,28}, M.~Beilicke\altaffilmark{5}, W.~Benbow\altaffilmark{1},
  S.~M.~Bradbury\altaffilmark{6}, J.~H.~Buckley\altaffilmark{5}, V.~Bugaev\altaffilmark{5},
  K.~Byrum\altaffilmark{7}, A.~Cannon\altaffilmark{8}, A.~Cesarini\altaffilmark{9},
  J.~L.~Christiansen\altaffilmark{10}, L.~Ciupik\altaffilmark{11},
  E.~Collins-Hughes\altaffilmark{8}, M.~P.~Connolly\altaffilmark{9}, W.~Cui\altaffilmark{12},
  C.~Duke\altaffilmark{13}, M.~Errando\altaffilmark{2}, A.~Falcone\altaffilmark{14},
  J.~P.~Finley\altaffilmark{12}, G.~Finnegan\altaffilmark{15}, L.~Fortson\altaffilmark{16},
  A.~Furniss\altaffilmark{4}, N.~Galante\altaffilmark{1}, D.~Gall\altaffilmark{12},
  S.~Godambe\altaffilmark{15}, S.~Griffin\altaffilmark{17}, J.~Grube\altaffilmark{11},
  R.~Guenette\altaffilmark{17}, G.~Gyuk\altaffilmark{11}, D.~Hanna\altaffilmark{17},
  J.~Holder\altaffilmark{18}, G.~Hughes\altaffilmark{19}, C.~M.~Hui\altaffilmark{15},
  T.~B.~Humensky\altaffilmark{20}, D.~J.~Jackson\altaffilmark{10}, P.~Kaaret\altaffilmark{21},
  N.~Karlsson\altaffilmark{16}, M.~Kertzman\altaffilmark{22}, D.~Kieda\altaffilmark{15},
  H.~Krawczynski\altaffilmark{5}, F.~Krennrich\altaffilmark{23}, M.~J.~Lang\altaffilmark{9},
  A.~S~Madhavan\altaffilmark{23}, G.~Maier\altaffilmark{19}, S.~McArthur\altaffilmark{5},
  A.~McCann\altaffilmark{17}, P.~Moriarty\altaffilmark{24}, M.~D.~Newbold\altaffilmark{15},
  R.~A.~Ong\altaffilmark{3}, M.~Orr\altaffilmark{23}, A.~N.~Otte\altaffilmark{4},
  N.~Park\altaffilmark{20}, J.~S.~Perkins\altaffilmark{1}, M.~Pohl\altaffilmark{25,19},
  H.~Prokoph\altaffilmark{19}, J.~Quinn\altaffilmark{8}, K.~Ragan\altaffilmark{17},
  L.~C.~Reyes\altaffilmark{20}, P.~T.~Reynolds\altaffilmark{26}, E.~Roache\altaffilmark{1},
  H.~J.~Rose\altaffilmark{6}, J.~Ruppel\altaffilmark{25}, D.~B.~Saxon\altaffilmark{18},
  M.~Schroedter\altaffilmark{23}, G.~H.~Sembroski\altaffilmark{12},
  G.~D.~\c{S}ent\"{u}rk\altaffilmark{27}, A.~W.~Smith\altaffilmark{7}, D.~Staszak\altaffilmark{17},
  S.~P.~Swordy\altaffilmark{29}, G.~Te\v{s}i\'{c}\altaffilmark{17}, M.~Theiling\altaffilmark{1},
  S.~Thibadeau\altaffilmark{5}, K.~Tsurusaki\altaffilmark{21}, A.~Varlotta\altaffilmark{12},
  V.~V.~Vassiliev\altaffilmark{3}, S.~Vincent\altaffilmark{15}, M.~Vivier\altaffilmark{18},
  S.~P.~Wakely\altaffilmark{20}, J.~E.~Ward\altaffilmark{8}, T.~C.~Weekes\altaffilmark{1},
  A.~Weinstein\altaffilmark{3}, T.~Weisgarber\altaffilmark{20}, D.~A.~Williams\altaffilmark{4},
  M.~Wood\altaffilmark{3} }

\altaffiltext{1}{Fred Lawrence Whipple Observatory, Harvard-Smithsonian Center for Astrophysics, Amado, AZ 85645, USA}
\altaffiltext{2}{Department of Physics and Astronomy, Barnard College, Columbia University, NY 10027, USA}
\altaffiltext{3}{Department of Physics and Astronomy, University of California, Los Angeles, CA 90095, USA}
\altaffiltext{4}{Santa Cruz Institute for Particle Physics and Department of Physics, University of California, Santa Cruz, CA 95064, USA}
\altaffiltext{5}{Department of Physics, Washington University, St. Louis, MO 63130, USA}
\altaffiltext{6}{School of Physics and Astronomy, University of Leeds, Leeds, LS2 9JT, UK}
\altaffiltext{7}{Argonne National Laboratory, 9700 S. Cass Avenue, Argonne, IL 60439, USA}
\altaffiltext{8}{School of Physics, University College Dublin, Belfield, Dublin 4, Ireland}
\altaffiltext{9}{School of Physics, National University of Ireland Galway, University Road, Galway, Ireland}
\altaffiltext{10}{Physics Department, California Polytechnic State University, San Luis Obispo, CA 94307, USA}
\altaffiltext{11}{Astronomy Department, Adler Planetarium and Astronomy Museum, Chicago, IL 60605, USA}
\altaffiltext{12}{Department of Physics, Purdue University, West Lafayette, IN 47907, USA }
\altaffiltext{13}{Department of Physics, Grinnell College, Grinnell, IA 50112-1690, USA}
\altaffiltext{14}{Department of Astronomy and Astrophysics, 525 Davey Lab, Pennsylvania State University, University Park, PA 16802, USA}
\altaffiltext{15}{Department of Physics and Astronomy, University of Utah, Salt Lake City, UT 84112, USA}
\altaffiltext{16}{School of Physics and Astronomy, University of Minnesota, Minneapolis, MN 55455, USA}
\altaffiltext{17}{Physics Department, McGill University, Montreal, QC H3A 2T8, Canada}
\altaffiltext{18}{Department of Physics and Astronomy and the Bartol Research Institute, University of Delaware, Newark, DE 19716, USA}
\altaffiltext{19}{DESY, Platanenallee 6, 15738 Zeuthen, Germany}
\altaffiltext{20}{Enrico Fermi Institute, University of Chicago, Chicago, IL 60637, USA}
\altaffiltext{21}{Department of Physics and Astronomy, University of Iowa, Van Allen Hall, Iowa City, IA 52242, USA}
\altaffiltext{22}{Department of Physics and Astronomy, DePauw University, Greencastle, IN 46135-0037, USA}
\altaffiltext{23}{Department of Physics and Astronomy, Iowa State University, Ames, IA 50011, USA}
\altaffiltext{24}{Department of Life and Physical Sciences, Galway-Mayo Institute of Technology, Dublin Road, Galway, Ireland}
\altaffiltext{25}{Institut f\"ur Physik und Astronomie, Universit\"at Potsdam, 14476 Potsdam-Golm,Germany}
\altaffiltext{26}{Department of Applied Physics and Instrumentation, Cork Institute of Technology, Bishopstown, Cork, Ireland}
\altaffiltext{27}{Columbia Astrophysics Laboratory, Columbia University, New York, NY 10027, USA}
\altaffiltext{28}{Author to whom correspondence may be addressed}
\altaffiltext{29}{Deceased}

\begin{abstract}
  We present the results of 16 {\it Swift}-triggered Gamma-ray burst (GRB) follow-up observations taken with the
  Very Energetic Radiation Imaging Telescope Array System (VERITAS) telescope array from 2007 January to 2009 June. The median energy threshold and response
  time of these observations were 260 GeV and 320 s, respectively. Observations had an average
  duration of 90 minutes. Each burst is analyzed independently in two modes: over the whole duration
  of the observations and again over a shorter timescale determined by the maximum VERITAS
  sensitivity to a burst with a $t^{-1.5}$ time profile.  This temporal model is characteristic of GRB
  afterglows with high-energy, long-lived emission that have been detected by the Large Area
  Telescope on board the {\it Fermi} satellite.
  No significant very high energy (VHE) gamma-ray emission was detected and upper limits
  above the VERITAS threshold energy are calculated. The VERITAS upper limits are corrected for
  gamma-ray extinction by the extragalactic background light and interpreted in the context of
  the keV emission detected by {\it Swift}. For some bursts the VHE emission 
  must have less power
  than the keV emission, placing constraints on inverse Compton models of VHE emission.
\end{abstract}


\section{Introduction}
Gamma-ray bursts (GRBs) have been an active area of study since their discovery in the late 1960s
\citep{1973ApJ...182L..85K}.  Observations of GRBs and their afterglows over the last decade are
generally consistent with the relativistic fireball framework
(e.g. \citealt{1999PhR...314..575P}). In this theoretical framework, prompt gamma-ray emission is
produced by internal shocks created by relativistic jets with varied Lorentz factors that originate
from a central engine. The afterglow emission arises from external shocks set up when outflowing
material interacts with the surrounding environment. Within this basic fireball framework there have
been a number of theories proposed that predict VHE photon production. A proposed physical
mechanism that produces VHE radiation in GRBs is inverse Compton (IC) scattering. By this mechanism
electrons accelerated by the burst's central engine up-scatter relatively soft photons from an
external photon field (external inverse Compton, \citealt{2005ApJ...618L..13B,
  2006ApJ...641L..89W}) or from a photon field generated by synchrotron emission from the electrons
themselves (synchrotron self Compton, SSC, \citealt{2001ApJ...559..110Z, 2001ApJ...556.1010W}).

External shocks may also produce VHE photons. If this is the case, measurements of the spectrum
above 10 GeV can directly constrain the medium density as well as the equipartition fraction of the
magnetic field \citep{2005ApJ...633.1018P} in the burst environment. VHE emission delayed by
$\sim$100 to $>$10000 s can be produced by the external forward shock \citep{1994MNRAS.269L..41M,
  2000ApJ...537..785D, 2008MNRAS.384.1483F}. GeV emission from electron synchrotron processes in the
forward shock has been predicted to be relatively bright \citep{2009MNRAS.396.1163Z} and it has been
proposed \citep{2009MNRAS.400L..75K} that such emission was detected by the Large Area Telescope
(LAT) on the {\it Fermi} satellite \citep{2009ApJ...697.1071A} in the bright gamma-ray burst
GRB\,080916C \citep{2009Sci...323.1688A}. In addition to the GeV synchrotron component, there may
also be SSC processes producing VHE photons in the forward shock \citep{2008MNRAS.385.1628P}. This
component is predicted to be less intense than the synchrotron component and therefore difficult to
detect with {\it Fermi}-LAT, but the very high energies and relatively late emission times (up to
several hours) make these photons prime candidates for detection by ground-based, imaging
atmospheric Cherenkov telescope (IACT) systems \citep{2009MNRAS.396.1163Z, 2009ApJ...703...60X}.

Yet another possible mechanism for generating delayed VHE photons from GRBs is IC scattering of
photons from X-ray flares.  The X-Ray Telescope (XRT) on board {\it Swift} has made it possible to
take detailed X-ray observations of fading GRB afterglows on a regular basis. In roughly half of
these observations, X-ray flare activity has been observed that takes place hundreds to thousands of
seconds after the initial gamma-ray signal \citep{2007ApJ...671.1903C}. It is predicted that VHE
photons could arise from the X-ray photons, produced by late-time central engine activity,
interacting with electrons accelerated at the forward shock. It is also possible that the X-ray
flares are produced by the forward shock itself and that VHE photons are consequently created
through the SSC process.  Simultaneous observations of X-ray and VHE afterglows can distinguish
between these two possibilities and can constrain the microphysics in the shocks themselves
\citep{2006ApJ...641L..89W}. While not expected to be a routine event, detection of VHE emission
from X-ray flares in GRBs by current generation IACTs (VERITAS, MAGIC, HESS) should be possible
under favorable conditions \citep{2008MNRAS.384.1483F, 2008A&A...489.1073G}. Recently, the {\it Fermi}-LAT 
detected hard-spectrum ($\Gamma=1.4$) high energy emission associated with late-time X-ray flaring activity in 
GRB\,100728A \citep{2011ApJ...734L..27A}. Finally, GRBs have
been advanced as a possible class of sources that generate ultra-high-energy cosmic rays 
\citep{2004ApJ...606..988W, 2008PhRvD..78b3005M, 2007ApJ...664..384D}. In hadronic or combined
leptonic/hadronic models, VHE gamma-rays are produced by the energetic leptons that are created from
cascades initiated by photopion production \citep{1998ApJ...499L.131B}.

There have been several attempts to observe VHE photon emission from GRBs using ground-based
facilities but, to date, no conclusive detections have been made. A possible detection of VHE gamma
rays associated with the BATSE-detected GRB\,970417A was reported \citep{2000ApJ...533L.119A} by the
Milagrito Collaboration but no redshift was determined and no other follow-up observations were
made.  Even though detection of VHE afterglow emission with IACTs is predicted to be possible,
observations by both previous \citep{1997ApJ...479..859C} and current-generation
\citep{2009A&A...495..505A, 2007ApJ...667..358A} observatories have yielded no significant
detections.

Presented here are the results from GRB observations made during an 18 month interval with the Very
Energetic Radiation Imaging Telescope Array System (VERITAS) between autumn 2007 and spring
2009. The sample is limited to well-localized bursts observed with at least three of the four
VERITAS telescopes.

\section{The VERITAS Array}
VERITAS is an array of four IACTs, each 12 m in diameter, located 1268 m a.s.l. at the Fred Lawrence
Whipple Observatory in southern Arizona, USA (31\degree $40' 30''$ N, 110\degree $57' 07''$ W). The
first telescope was completed in the spring of 2005 and the full, four-telescope array began routine
observations in the autumn of 2007. The first telescope was installed at a temporary location as a
prototype instrument and in the summer of 2009 it was moved to a new location in the array to make
the distance between telescopes more uniform and consequently improve the sensitivity of the system
\citep{2009arXiv0912.3841P}. The observations presented here were taken with the old array
configuration with at least three telescopes in the array operational. Each of the telescopes is of
Davies-Cotton design and is equipped with an imaging camera consisting of 499 photomultiplier tubes
(PMTs) at the focus, 12 m from the center of the reflector.  The angular spacing of the PMTs is
approximately $0\degree\!\!.15$ resulting in a field of view of $3\degree\!\!.5$ in diameter. Each PMT has a
Winston cone mounted in front of the cathode to reduce the dead space between pixels and to increase
the light collection efficiency.

The VERITAS array uses a three-level trigger system that greatly reduces the number of background
events. The first level of the trigger system is at the pixel, i.e. PMT, level where the signal from
each PMT is fed to a programmable constant fraction discriminator with a threshold of 4--5
photoelectrons. The second trigger level, the camera/telescope trigger, consists of a pattern
trigger that requires at least three adjacent pixels satisfying the first level trigger within a
$\sim 7$ ns coincidence window. Finally, an array-level trigger is satisfied if at least two of the
four telescopes in the array are triggered within 100 ns of one another, after correcting for
time-of-flight differences. Once the array is triggered, the signals, which are continuously
digitized for each PMT using 500 mega-samples per second flash analog to digital converters (FADCs),
are read out and stored to disk. The array has an effective area of $\sim 10^{3} \,\rm{m}^{2}$ to
$\sim 10^{5} \,\rm{m}^{2}$ and an energy resolution of $15$ - $20 \%$ over the $100\, \rm{GeV}$ -
$30\,\rm{TeV}$ energy range. The single event angular resolution (68\% containment) is better than
$0^{\circ}.14$. A more comprehensive description of the VERITAS array can be found in
\citet{2006APh....25..391H}.

\section{Gamma-ray burst observations}

\begin{table*}
  \caption{Details of 16 GRBs observed by VERITAS}
  \label{tab:swiftbursts}
  \begin{centering}
  \begin{tabular*}{\textwidth}{@{\extracolsep{\fill}}ccccccccc}
    \hline \hline
    GRB & {\it Swift} trigger & T$_{90}$(s)$^{\alpha}$ & 
    Fluence ($10^{-7}$erg cm$^{-2}$)$^{\beta}$ & T$_{\rm trig}^{\gamma}$ & RA & Dec & Error & $z$ \\
    \hline
    070223 & $261664$ & $89$ & $17$ & 01:15:00 & $10{\rm\hour}13{\rm\minute}48{\rm\second}\!\!.39$ 
    & $+43\degree08'00.70''$ & $0.30''$ & ... \\
    070419A & $276205$ & $116$ & $5.6$ & 09:59:26 & $12{\rm\hour}10{\rm\minute}58{\rm\second}\!\!.83$ 
    & $+39\degree55'34.06''$ & $0.15''$ & $0.97^{\delta}$ \\
    070521 & $279935$ & $37.9$ & $80$ & 06:51:10 & $16{\rm\hour}10{\rm\minute}38{\rm\second}\!\!.59$ 
    & $+30\degree15'21.96''$ & $1.70''$ & $0.553?^{\epsilon}$ \\
    070612B & $282073$ & $13.5$ & $17$ & 06:21:17 & $17{\rm\hour}26{\rm\minute}54{\rm\second}\!\!.49$ 
    & $-08\degree45'06.3''$ & $4.0''$ & ... \\
    071020 & $294835$ & $4.2$ & $23$ & 07:02:26 & $07{\rm\hour}58{\rm\minute}39{\rm\second}\!\!.78$ 
    & $+32\degree51'40.4''$ & $0.250''$ & $2.145^{\zeta}$ \\
    080129 & $301981$ & $48$ & $8.9$ & 06:06:45 & $07{\rm\hour}01{\rm\minute}08{\rm\second}\!\!.20$ 
    & $-07\degree50'46.3''$ & $0.3''$ & ... \\
    080310 & $305288$ & $365$ & $23$ & 08:37:58 & $14{\rm\hour}40{\rm\minute}13{\rm\second}\!\!.80$ 
    & $-00\degree10'29.60''$ & $0.6''$ & $2.43^{\eta}$ \\
    080330 & $308041$ & $61$ & $3.4$ & 03:41:16 & $11{\rm\hour}17{\rm\minute}04{\rm\second}\!\!.50$ 
    & $+30\degree37'23.53''$ & $0.7''$ & $1.51^{\theta}$ \\ 
    080409 & $308812$ & $20.2$ & $6.1$ & 01:22:57 & $05{\rm\hour}37{\rm\minute}19{\rm\second}\!\!.14$ 
    & $+05\degree05'05.4''$ & $2.0''$ & ... \\
    080604 & $313116$ & $82$ & $8.0$ & 07:27:01 & $15{\rm\hour}47{\rm\minute}51{\rm\second}\!\!.70$ 
    & $+20\degree33'28.1''$ & $0.5''$ & $1.416^{\iota}$ \\ 
    080607 & $313417$ & $79$ & $240$ & 06:07:27 & $12{\rm\hour}59{\rm\minute}47{\rm\second}\!\!.24$ 
    & $+15\degree55'08.74''$ & $0.5''$ & $3.036^{\kappa}$ \\
    081024A & $332516$ & $1.8$ & $1.2$ & 05:54:21 & $01{\rm\hour}51{\rm\minute}29{\rm\second}\!\!.71$ 
    & $+61\degree19'53.04''$ & $1.9''$ & ... \\
    090102 & $338895$ & $27$ & $68$ & 02:55:45 & $08{\rm\hour}32{\rm\minute}58{\rm\second}\!\!.54$ 
    & $+33\degree06'51.10''$ & $0.5''$ & $1.55^{\lambda}$ \\
    090418A & $349510$ & $56$ & $46$ & 11:07:40 & $17{\rm\hour}57{\rm\minute}15{\rm\second}\!\!.17$ 
    & $+33\degree24'21.1''$ & $0.5''$ & $1.608^{\mu}$ \\
    090429B & $350854$ & $5.5$ & $3.1$ & 05:30:03 & $14{\rm\hour}02{\rm\minute}40{\rm\second}\!\!.10$ 
    & $+32\degree10'14.6''$ & $1.8''$ & ... \\
    090515 & $352108$ & $0.036$ & $0.04$ & 04:45:09 & $10{\rm\hour}56{\rm\minute}36{\rm\second}\!\!.11$ 
    & $+14\degree26'30.3''$ & $2.7''$ & ... \\
    \hline\hline
  \end{tabular*}
\end{centering}
All information was taken from GCN circulars (http://gcn.gsfc.nasa.gov/gcn3\_archive.html).
$^{\alpha}$ Duration over which 90\% of the emission in the 15--350 keV energy band occurs, as
measured by the {\it Swift} Burst Alert Telescope (BAT).  $^{\beta}$ 15--150 keV fluence, as measured by the
{\it Swift}-BAT.  $^{\gamma}$ UT time of the GRB trigger determined by the {\it
  Swift}-BAT.  $^{\delta}$\citet{Cenko:2007p3249}. $^{\epsilon}$\citet{Hattori:2007p3267}.
$^{\zeta}$\citet{Jakobsson:2007p3268}. $^{\eta}$\citet{Prochaska:2008p3290}.
$^{\theta}$\citet{Malesani:2008p3305}. $^{\iota}$\citet{Wiersema:2008p3333}.
$^{\kappa}$\citet{Prochaska:2008p3342}. $^{\lambda}$\citet{deUgartePostigo:2009p3363}.
$^{\mu}$\citet{Chornock:2009p3395}.
\end{table*}    

\begin{figure*}[htp]
  \begin{center}
    \scalebox{1}{
      \plotone{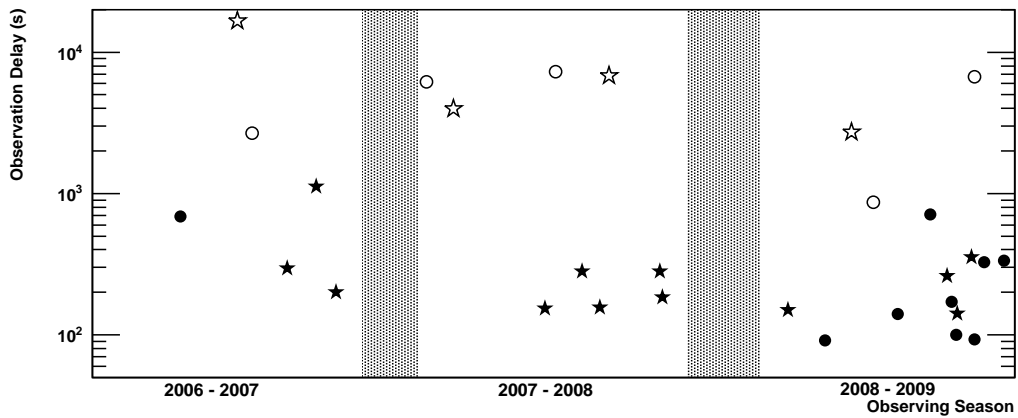}
    }
  \end{center}
  \caption{Delay from the start of the burst to the beginning of VERITAS observations for all GRBs
    with VERITAS data. The open symbols correspond to observations that were delayed due to
    constraints such as the burst occurring during daylight or below the horizon. Filled symbols are
    unconstrained observation delays and are primarily determined by the time it takes the
    telescopes to slew to the burst. The 16 stars correspond to the GRBs discussed in this
    paper. The shaded regions indicate the annual shutdown of the array due to the summer monsoons.}
  \label{fig:GRBObservationDelays}
\end{figure*}

GRB observations take priority over all others in the VERITAS observing plan. To facilitate rapid
follow-up observations of GRBs detected by satellites, VERITAS control computers are set to receive
notices from the GRB Coordinates Network (GCN)\footnote{http://gcn.gsfc.nasa.gov} over a socket
connection through the TCP/IP protocol. Once the GCN notice is parsed by the control computer, an
audible alarm notifies the observers on duty that a GRB has occurred.  The coordinates of the burst
are loaded into the telescope tracking software and the observers are notified to stop current
observations and to begin slewing the telescopes to the GRB position, subject to observational
constraints such as the Moon and horizon. Currently, the telescopes are capable of simultaneously
slewing at a rate of 1\degree s$^{-1}$ in both elevation and azimuth. Figure \ref{fig:GRBObservationDelays} shows the observation delays
for all GRBs with VERITAS data over a three-year period. The delay between the satellite
trigger and the beginning of GRB observations is usually less than 300 s if the burst is immediately 
observable, and in several cases this delay is less
than 100 s. The dominant contribution to the observation delay is the time it takes the telescopes
to slew to the source position. 

If the GRB is sufficiently well-localized, as is the case with the bursts presented here, VERITAS
observations continue for up to 3 hr after the GRB satellite trigger, again subject to
observing constraints. The transition from the prompt to the afterglow phase of a GRB, which can
occur hundreds to thousands of seconds after the initial burst, is often accompanied by X-ray flares
\citep{2007ApJ...671.1921F}. These flares can be very bright and may be associated with extended
activity from the GRB central engine \citep{2005Sci...309.1833B} or be from delayed external shocks
that could produce a relatively large flux of gamma rays in the $\sim100$ GeV energy range. For
GRBs, the VERITAS strategy of rapid follow-up observations that continue for several hours allows
for good temporal coverage of X-ray flare phenomena. Even in the absence of flare activity, it is
suggested that a significant flux of high-energy photons from IC processes associated with the GRB
afterglow may extend to more than 10 ks after the beginning of the GRB prompt emission
\citep{2008A&A...489.1073G} and so an observation window of several hours is warranted.

During the period beginning 2007 January and ending 2009 June, VERITAS took follow-up observations
of 29 GRBs. Nine of these bursts were detected only by the Gamma-ray Burst Monitor (GBM) on board
the {\it Fermi} satellite and the errors on the localizations were larger than the VERITAS FOV. 
Analysis of these bursts will be presented in a future publication. The remaining sample of
20 well-localized bursts (19 detected by the {\it Swift} satellite and 1 by the {\it INTEGRAL}
spacecraft) is reduced to 16 after applying cuts on the hardware status of the array and on the
burst elevation (minimum elevation for GRB observations considered in this analysis is 30\degree). 
Table \ref{tab:swiftbursts} lists the general properties of these 16 bursts. The
VERITAS observations of GRBs presented here took place during good weather and under dark skies or
low-moonlight conditions.

The data were collected in runs with nominal durations of 20 minutes with roughly 30 seconds
of dead time between runs. At the beginning of each run the best source localization to arrive via
the GCN socket connection is used as the target for the duration of that run. Twelve of the bursts
were observed in ``wobble mode'' in which the source is displaced some angular distance away from
the center of the camera, allowing simultaneous observation and background estimation
\citep{2007A&A...466.1219B}. For the GRB observations presented here, the wobble offset was
$0\degree\!\!.5$.  
In the cases of GRB\,070419A, GRB\,070521, GRB\,070612B, and GRB\,080604, observations
were taken in a tracking mode in which the source is placed at the center of the camera. Historically, 
GRB observations were taken in tracking mode but wobble mode 
is now the default method of observation with VERITAS and all GRB observations are currently
taken in this fashion. The use of the tracking mode does offer a marginal increase in ``raw'' sensitivity over 
the wobble mode but with a significant increase in the uncertainty of the background. 

\section{Data analysis}
The data taken on the 16 GRBs were analyzed using the standard VERITAS analysis suite
\citep{2008ICRC....3.1385C}. The charge in each FADC trace is determined by summing the samples over
an appropriately-placed 14 ns-wide integration window. The integrated signal from each pixel in the
camera results in an image of the air shower at the camera plane. The shower image is cleaned by
eliminating any pixel with a signal of less than five standard deviations above its pedestal value,
that is, a signal less than five times the standard deviation from the average FADC measurement when
no Cherenkov signal is present. Any pixel that registers a signal of at least two and a half
standard deviations above its pedestal is also retained provided it is adjacent to at least one of
the pixels that exceeds five standard deviations. The cleaned images are then parameterized using
the Hillas moment analysis \citep{1985ICRC....3..445H}. Before performing a full event
reconstruction, images with less than five pixels surviving the image cleaning or with an image
centroid more than $1\degree\!\!.43$ from the camera center are removed from the analysis. A cut on the
integrated charge in each image is made at $\sim75$ ($\sim38$) photoelectrons for the
standard-source (soft-source) analysis. For GRBs, the standard-source analysis is optimized for a
weak Crab-like source (3\% Crab flux with a spectral index, $\Gamma=2.5$), while the soft-source
analysis gives a reduced energy threshold and assumes a $\Gamma=3.5$ spectrum. While the spectral
characteristics of GRBs are unknown at the highest energies, the standard analysis spectral index of
2.5 was selected based on the average high-energy spectral index, $\beta$ observed by BATSE
\citep{Kaneko:2006p3159}. Since it is expected that the extragalactic background light (EBL) will significantly soften the intrinsic
GRB spectrum, the soft source analysis was optimized to the softer assumed spectral index of 3.5. It
should be noted that although the analysis is optimized for a specific spectral index and source
intensity, this does not preclude the detection of sources with characteristics significantly
different than those assumed.

At this stage, any event with images in fewer than two telescopes is rejected because stereo
reconstruction is not possible. Furthermore, any event with images in only the two telescopes with
the smallest separation is removed as the proximity of these two telescopes ($\sim35$ m) in the old
array configuration produced less reliable event reconstruction and an increased background rate
that resulted in decreased sensitivity. After event reconstruction, the rejection of background
events, which are due largely to cosmic rays, is accomplished by comparing the length and width
parameters of shower images with those predicted by Monte Carlo simulations of gamma ray-initiated
air showers \citep{2006APh....25..380K}. Finally, a cut on the arrival direction of the gamma ray of
$\theta < 0\degree\!\!.13$ ($\theta < 0\degree\!\!.14$) for the standard (soft) analysis is applied,
where $\theta$ is the angular distance in the FOV from the reconstructed arrival direction
of the event to the putative source location. For all bursts presented here, the uncertainty in the
GRB position (in all cases $<4''$) is negligible compared to the angular distance cut on arrival
direction.

Twelve of the sixteen bursts were observed in ``wobble'' mode and the estimation of the background
in the signal region is made using the reflected region technique \citep{2001A&A...370..112A}. In
the cases of GRB\,070419A, GRB\,070521, GRB\,070612B, and GRB\,080604, the observations were made
with the GRB positions at the center of the fields of view of the telescopes and a reflected region
background estimation is not possible. For these observations the ring-background estimation method
\citep{2007A&A...466.1219B} is employed instead. The significance of the gamma-ray excess in the
signal region is then computed using Equation 17 in \citet{1983ApJ...272..317L}. 

If there is no significant gamma-ray excess detected (i.e., the excess in the signal region is less
than five standard deviations above the background region), the 99\% confidence level upper limit on
the number of signal photons is calculated using the frequentist method of
\citet{2005NIMPA.551..493R}. From this number, the corresponding upper limit on the integral photon
flux above the threshold energy is computed. The energy threshold is defined as the energy at which
the product of the detector effective area and assumed source spectrum is maximized. The effective
area, and consequently the threshold energy, of VERITAS is strongly dependent on the elevation of
the source being observed. As the elevation of the observation decreases, the column density of the
atmosphere increases. This results in a gamma ray of some given energy producing a lower Cherenkov
photon density at ground level, which increases the energy threshold of detection. However, because
the effective area of the instrument is non-zero below the threshold energy defined in this way,
gamma rays in this energy range are detectable. For all of the VERITAS data analyzed, a secondary
analysis was done using an independent software package and the results obtained are compatible with
those presented here.

A search for VHE emission is performed over the entire duration of the VERITAS observations as well
as over a shorter timescale that optimizes the sensitivity of VERITAS to a source with a flux that
decays as a power-law in time. The {\it Fermi}-LAT has detected more than a dozen GRBs
with emission above 100 MeV. This high-energy emission is seen to persist after the flux in the GBM
band has ceased and shows weak spectral evolution with a spectral index between the $\alpha$ and
$\beta$ indices of the Band function fit to the GBM data \citep{Ghisellini:2010p2000}. The temporal
behavior of the brightest four {\it Fermi}-LAT detected bursts: GRB\,080916C
\citep{2009Sci...323.1688A}, GRB\,090510 \citep{DePasquale:2010p2521}, GRB\,090902B
\citep{2009ApJ...706L.138A}, and GRB\,090926A \citep{2011arXiv1101.2082F}, show a common
$\frac{dN}{dE} \sim t^{-\Delta}$ decay, where $1.2 < \Delta < 1.7$ in the observer
frame. If it is assumed that the temporal and spectral characteristics of a GRB detected by the {\it
  Fermi}-LAT extend to the VHE energy range, the observed power law temporal decay of the high
energy emission consequently defines an optimal duration over which the search for VHE emission is
maximally sensitive. This optimal duration is determined solely by the high-energy temporal power
law index of the GRB, the delay from the GRB trigger time ($T_{\rm trig}$) to the beginning of VERITAS GRB
observations, and by, to a lesser extent, the observational backgrounds. For a VERITAS observation
beginning 100 s after the GRB $T_{\rm trig}$, the observation window that gives maximum sensitivity is
$\sim 2$ -- $5$ minutes for GRBs similar to the brightest LAT-detected bursts. For bursts with
unknown high-energy behavior, the determination of an optimal time window for VHE observations is
not straightforward. However, the maximum sensitivity of a VHE instrument such as VERITAS to a GRB
with a power-law decay in time is likely to be on the order of a few minutes.

\begin{figure*}[htp]
  \begin{center}
    \scalebox{1}{
      \plotone{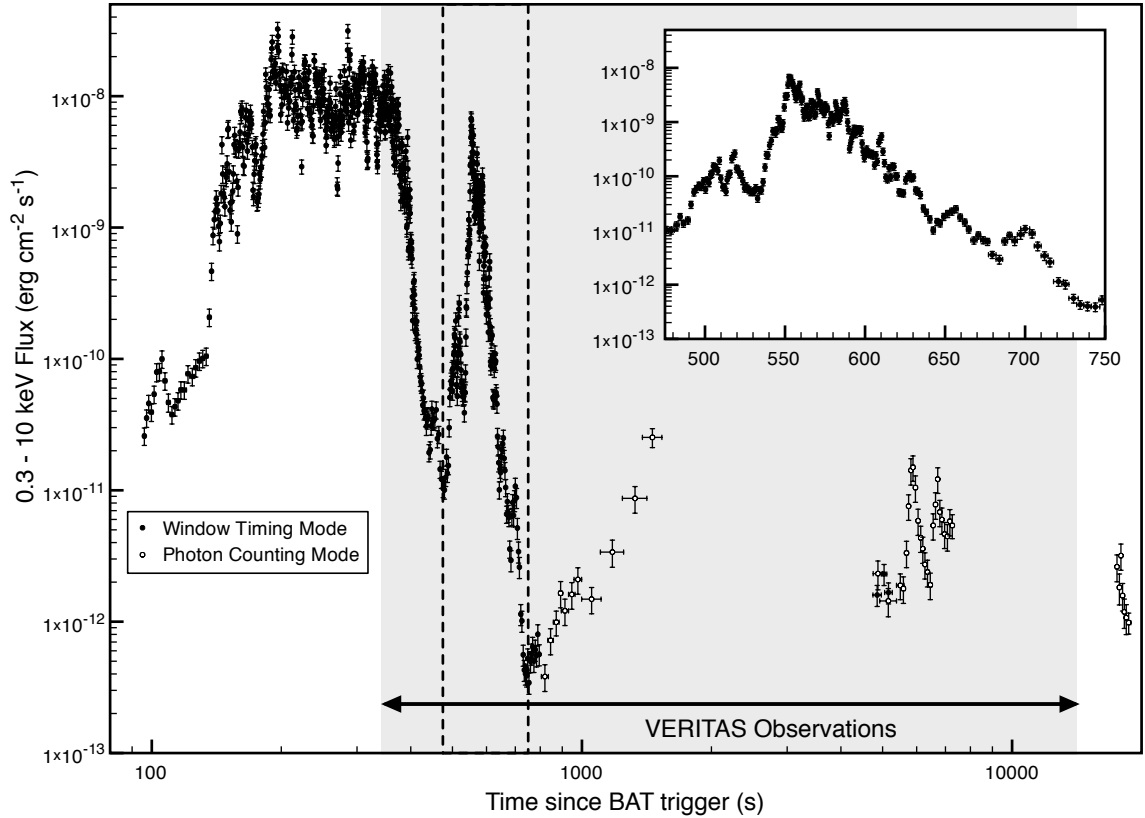}
    }
  \end{center}
  \caption{VERITAS observation window of GRB\,080310 superimposed on the {\it Swift}-XRT light curve
    \citep{Evans:2007p2698,Evans:2009p2640}. The inset shows the structure of the X-ray flare
    (dashed lines) and is the time window over which the search for VHE emission was performed. No
    significant excess of VHE gamma rays coincident with the X-ray flare (475 s $< t-T_{\rm trig}<$ 750 s)
    was found.}
  \label{fig:GRB080310XRTLC}
\end{figure*}

In the case of GRB\,080310, the {\it Swift}-XRT detected a large X-ray flare beginning $\sim 475$ s
after the beginning of the burst as measured by the {\it Swift}-BAT. VERITAS was on target 342 s
after $T_{\rm trig}$ for this burst and observed throughout the X-ray flare. Figure
\ref{fig:GRB080310XRTLC} shows the VERITAS observing window for this burst relative to the XRT
light curve \citep{Evans:2007p2698,Evans:2009p2640}. A search for VHE emission is made coincident
with the X-ray flare. 

\section{Results}
\begin{figure*}[htp]
  \begin{center}
    \subfigure[Standard source analysis]{\label{fig:edge-a}
      \scalebox{0.47}{
        \plotone{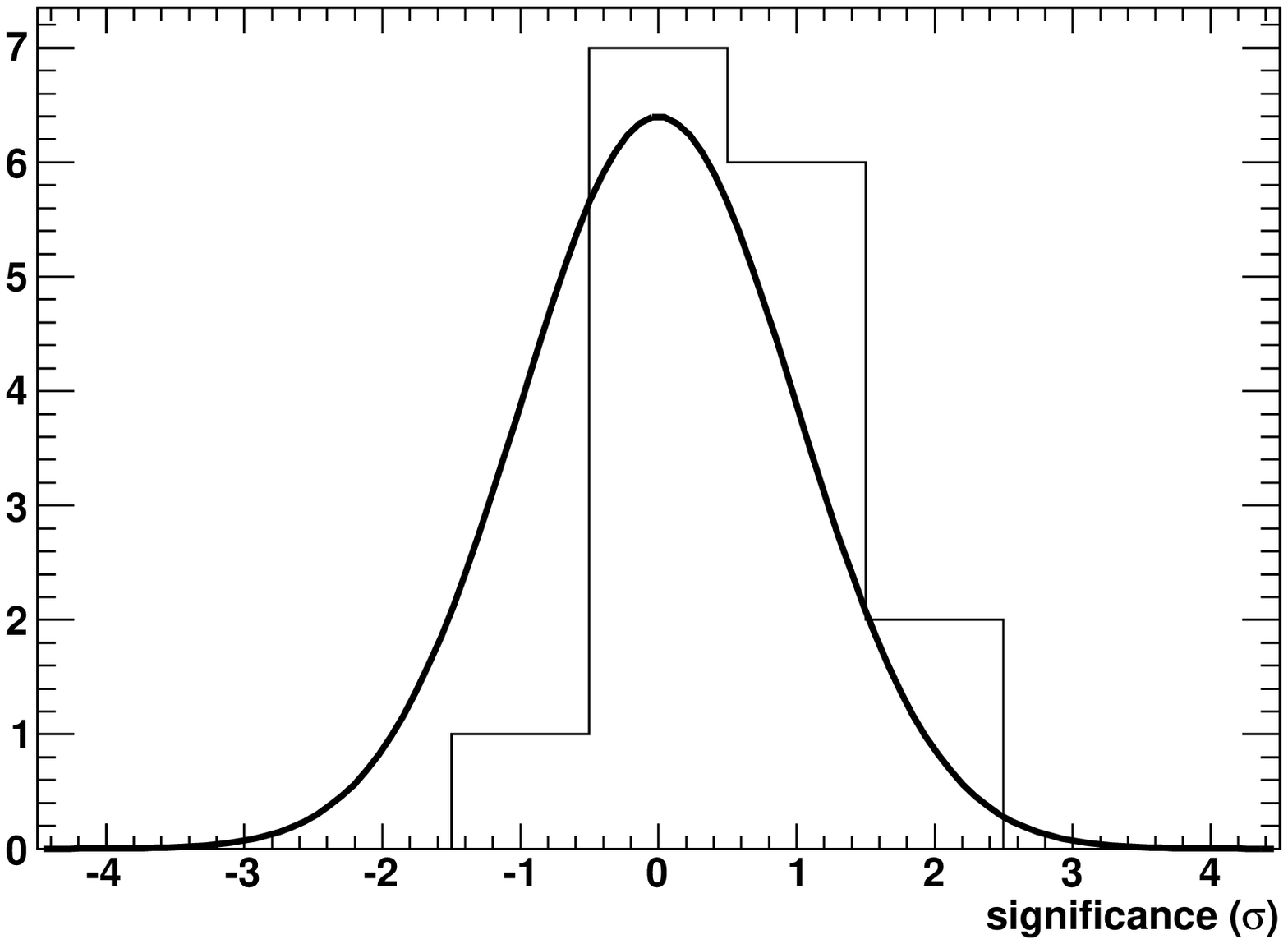}}
    }
    \subfigure[Soft source analysis]{\label{fig:edge-b}
      \scalebox{0.47}{
        \plotone{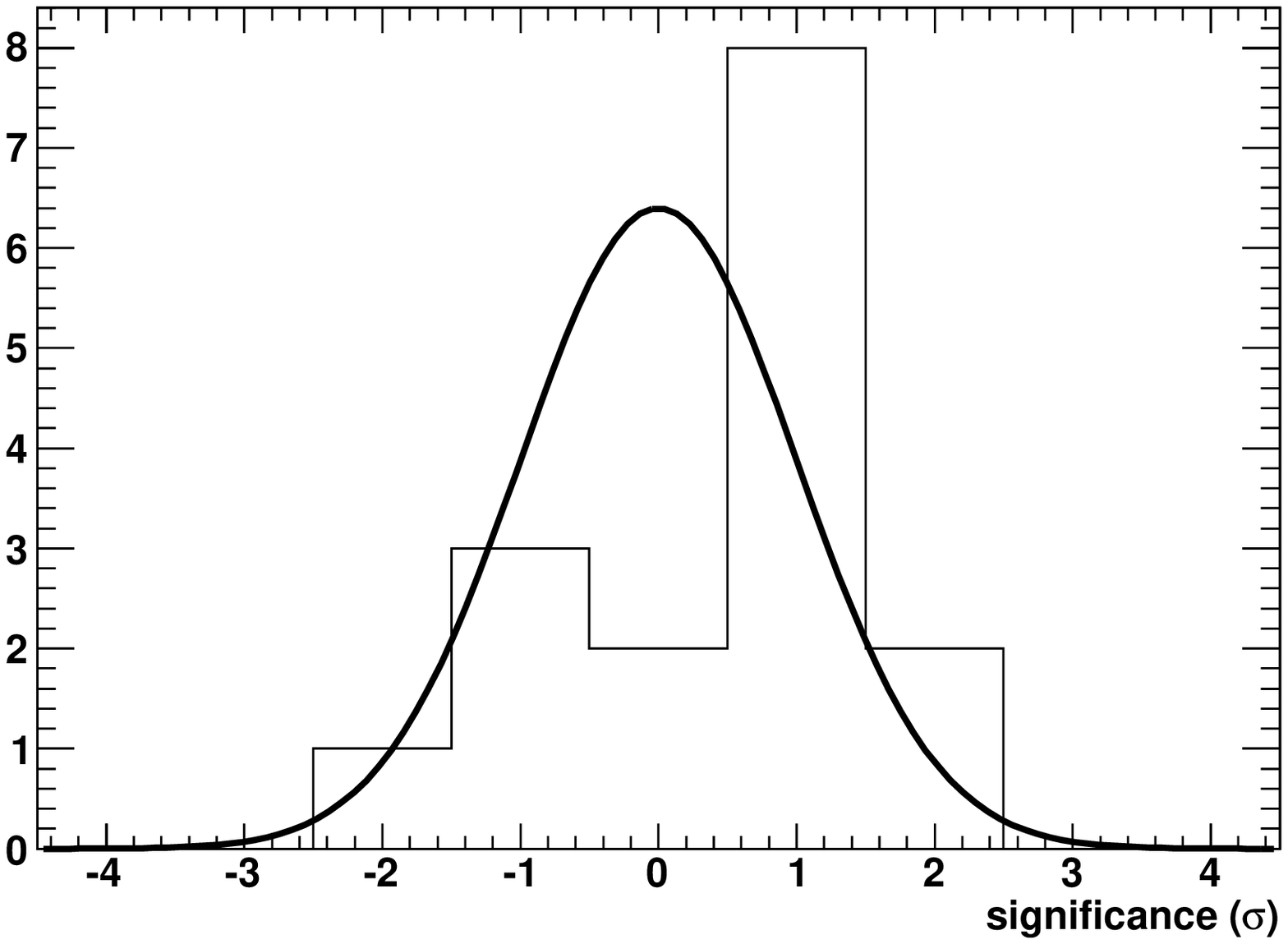}}
    }
  \end{center}
  \caption{Significance histograms of the 16 GRBs in this sample for both standard and soft source
    analyses. Included in the figures is the normalized Gaussian distribution of mean zero and
    variance one that the significance histograms should follow if no signal is present. The GRB
    significances are consistent with having been drawn from the aforementioned Gaussian
    distribution.}
  \label{fig:sigdistro}
\end{figure*}

An analysis of VERITAS data associated with the 16 GRB positions listed in Table
\ref{tab:swiftbursts} shows no significant excess of VHE gamma-ray events for any GRB over the
entire duration of VERITAS observations. Table \ref{tab:veritasbursts} summarizes the details and
results of the VERITAS GRB observations for the sample of GRBs described in Table
\ref{tab:swiftbursts}. The significance distributions for both the standard and soft source analyses
are shown in Figure \ref{fig:sigdistro}. The sensitivity of the VERITAS array, and the small
observation delays with respect to the GRB T$_{\rm trig}$ (half of the burst observations had delays of less
than 5 minutes) combine to give some of the most constraining limits on VHE gamma-ray emission
from GRB afterglows.

\begin{table*}
  \begin{small}
  \caption{VERITAS Observations of Gamma-Ray Bursts}
  \label{tab:veritasbursts}
  \begin{centering}
  \begin{tabular*}{\textwidth}{@{\extracolsep{\fill}}cccc|ccc|ccc}
    \hline \hline
    \multicolumn{4}{c|}{} & \multicolumn{3}{c|}{Standard Source Analysis} & 
    \multicolumn{3}{c}{Soft Source Analysis} \\
    GRB & T$_{\rm{delay}}$ (s)$^{\alpha}$ & T$_{\rm{obs}}$ (min)$^{\beta}$ & Elev. Range (\degree)$^{\gamma}$ & 
    $E_{\rm{th}}$ (GeV)$^{\delta}$ & $\sigma^{\zeta}$ & Upper Limit & 
    $E_{\rm{th}}$ (GeV)$^{\delta}$ & $\sigma^{\zeta}$ & Upper Limit \\
    \hline
    070223 & $1.7 \times 10^{4}$ & $74.1$ & 67--78 & $220$ & $1.3$ & $9.5 \times 10^{-12}$ & $150$ & $0.8$ & $2.0 \times 10^{-11}$ \\
    070419A & $295$ & $37.7$ & 32--36 & $610$ & $-0.1$ & $8.1 \times 10^{-12}$ & $420$ & $-1.0$ & $1.0 \times 10^{-11}$ \\
    070521 & $1118$ & $75.4$ & 63--88 & $190$ & $0.1$ & $4.6 \times 10^{-12}$ & $120$ & $-0.3$ & $9.6 \times 10^{-11}$ \\
    070612B & $201$ & $131.9$ & 46--50 & $380$ & $0.6$ & $2.5 \times 10^{-12}$ & $230$ & $0.6$ & $7.1 \times 10^{-12}$ \\
    071020 & $5259$ & $73.5$ & 30--43 & $570$ & $1.8$ & $1.7 \times 10^{-11}$ & $330$ & $0.5$ & $2.6 \times 10^{-11}$ \\
    080129 & $1456$ & $31.4$ & 47--50 & $370$ & $1.2$ & $7.7 \times 10^{-12}$ & $220$ & $1.4$ & $1.2 \times 10^{-11}$ \\
    080310 & $342$ & $198.0$ & 48--58 & $270$ & $0.2$ & $2.2 \times 10^{-12}$ & $170$ & $1.8$ & $7.3 \times 10^{-12}$ \\
    080330 & $156$ & $107.8$ & 64--88 & $180$ & $0.2$ & $4.0 \times 10^{-12}$ & $120$ & $-0.7$ & $6.3 \times 10^{-12}$ \\
    080409 & $6829$ & $19.0$ & 31--35 & $1300$ & $0.1$ & $5.3 \times 10^{-11}$ & $720$ & $-0.7$ & $3.8 \times 10^{-11}$ \\
    080604 & $281$ & $151.8$ & 33--70 & $250$ & $1.1$ & $4.7 \times 10^{-12}$ & $160$ & $0.9$ & $1.2 \times 10^{-11}$ \\
    080607 & $184$ & $56.0$ & 32--46 & $400$ & $1.5$ & $1.6 \times 10^{-11}$ & $310$ & $1.1$ & $2.4 \times 10^{-11}$ \\
    081024A & $150$ & $161.2$ & 55--60 & $310$ & $-1.5$ & $1.5 \times 10^{-12}$ & $190$ & $-2.0$ & $2.2 \times 10^{-12}$ \\
    090102 & $5344$ & $83.1$ & 33--48 & $400$ & $-0.1$ & $8.4 \times 10^{-12}$ & $230$ & $-0.3$ & $1.8 \times 10^{-11}$ \\
    090418A & $261$ & $30.4$ & 86--88 & $190$ & $1.0$ & $1.0 \times 10^{-11}$ & $120$ & $1.7$ & $3.0 \times 10^{-11}$ \\
    090429B & $141$ & $158.8$ & 70--88 & $180$ & $1.1$ & $3.9 \times 10^{-12}$ & $120$ & $1.0$ & $9.6 \times 10^{-12}$ \\
    090515 & $356$ & $78.8$ & 37--57 & $340$ & $0.1$ & $6.3 \times 10^{-12}$ & $220$ & $1.2$ & $2.5 \times 10^{-11}$ \\
    \hline\hline
  \end{tabular*}
\end{centering}
\end{small}
Upper limits are given at the 99\% confidence level in terms of $\nu F_{\nu}$ at $E_{\rm{th}}$,
assuming the spectral indices of 2.5 and 3.5 for the standard source and soft source analysis,
respectively, in units of erg cm$^{-2}$ s$^{-1}$. $^{\alpha}$Time between the GRB
trigger time ($T_{\rm trig}$) and the beginning of VERITAS GRB observation. $^{\beta}$Duration of VERITAS
observation. $^{\gamma}$Elevation range of the VERITAS observation. $^{\delta}$The VERITAS energy
threshold. $^{\zeta}$Statistical significance (standard deviations) of signal counts observed by
VERITAS at the GRB position.
\end{table*}

The VHE photon fluxes from objects at cosmological distances are attenuated due to photon absorption
by the EBL. Of the sixteen bursts for which results are presented here, nine had redshifts
determined by optical followup observations. For these bursts, a limit on the intrinsic photon flux
of the GRB can be set if one assumes a model of the EBL. For all calculations requiring a model of
the EBL, we use the model described in \citet{2009MNRAS.399.1694G}. To determine the factor by which
the upper limits in Table \ref{tab:veritasbursts} increase due to effects of the EBL, one must
calculate the effective attenuation of VHE photons over the VERITAS waveband, taking into account
the spectral response of the instrument.  For each GRB observation, the effective area of VERITAS is
multiplied by the assumed intrinsic spectrum of the burst, which we take to be $\Gamma=2.5$.  The total flux is then calculated by
integrating the intrinsic differential flux of the GRB multiplied by the effective area of VERITAS,
over all energies at which the product is non-negligible. This process is repeated, substituting an
EBL-attenuated burst spectrum for the intrinsic burst spectrum.  The ratio of the total photon flux
obtained using the intrinsic burst spectrum to the total photon flux obtained using the
EBL-attenuated burst spectrum gives the attenuation factor for that particular GRB observation. For the EBL-corrected upper limits obtained 
using the soft-source analysis, there is an extra correction factor to account for the assumed intrinsic burst spectrum
($\Gamma=2.5$) relative to the limits obtained in Table \ref{tab:veritasbursts} which assumes a softer observed spectrum ($\Gamma=3.5$).  The
attenuation factors and redshift-corrected upper limits for GRBs with known redshift are shown in
Table \ref{tab:redshiftTable}.  Not surprisingly, the attenuation depends strongly on both the
redshift and the energy threshold for a particular observation, but under good observing conditions,
specifically at small zenith angles, reasonable sensitivity out to $z \sim 2$ is attainable with
VERITAS.

\begin{table*}
  \begin{small}
  \caption{Redshift-corrected VERITAS Upper Limits on VHE Emission from Nine {\it Swift}-detected GRBs} 
  \label{tab:redshiftTable}
  \begin{centering}
  \begin{tabular*}{\textwidth}{@{\extracolsep{\fill}}cccccc}
    \hline \hline
    GRB & Redshift & Attenuation & Standard Source Analysis & 
    Soft Source Analysis \\
     & ($z$) & Factor & Upper Limit & Upper Limit \\
    \hline
    070419A & $0.97$ & $1.5 \times 10^{-4}$ & $5.4 \times 10^{-8}$ & $2.8 \times 10^{-8}$ \\
    070521 & $0.553$ & $0.2$ & $2.1 \times 10^{-11}$ & $2.9 \times 10^{-11}$ \\
    071020 & $2.145$ & $1.2 \times 10^{-8}$ & $1.5 \times 10^{-3}$ & $7.0 \times 10^{-4}$ \\
    080310 & $2.43$ & $3.1 \times 10^{-4}$ & $7.0 \times 10^{-9}$ & $1.4 \times 10^{-8}$ \\
    080330 & $1.51$ & $0.027$ & $1.5 \times 10^{-10}$ & $1.2 \times 10^{-10}$ \\
    080604 & $1.4$ & $4.7 \times 10^{-3}$ & $1.0 \times 10^{-9}$ & $9.9 \times 10^{-10}$ \\
    080607 & $3.036$ & $1.6 \times 10^{-7}$ & $1.1 \times 10^{-4}$ & $6.8 \times 10^{-5}$ \\
    090102 & $1.55$ & $7.1 \times 10^{-5}$ & $1.2 \times 10^{-7}$ & $8.1 \times 10^{-8}$ \\
    090418A & $1.608$ & $0.03$ & $3.1 \times 10^{-10}$ & $6.0 \times 10^{-10}$ \\
    \hline\hline
  \end{tabular*}
\end{centering}
\end{small}
Upper limit and threshold energy ($E_{\rm{th}}$) of each GRB defined as in Table 2. The attenuation factor is explained in the text. 
\end{table*}

\begin{figure*}[htp]
  \begin{center}
    \subfigure[Standard source analysis]{\label{fig:edge-a}
      \scalebox{0.47}{
        \plotone{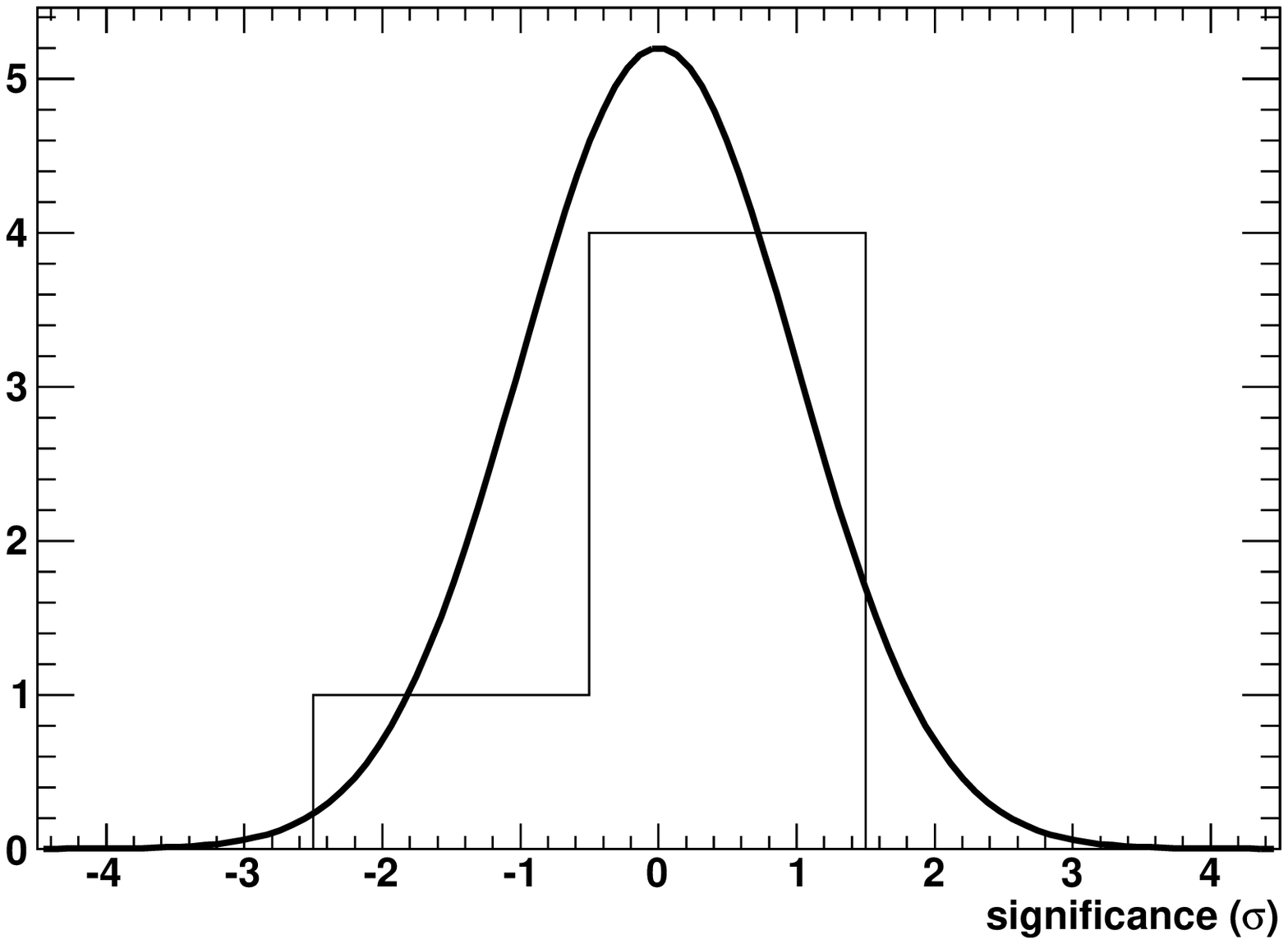}}
    }
    \subfigure[Soft source analysis]{\label{fig:edge-b}
      \scalebox{0.47}{
        \plotone{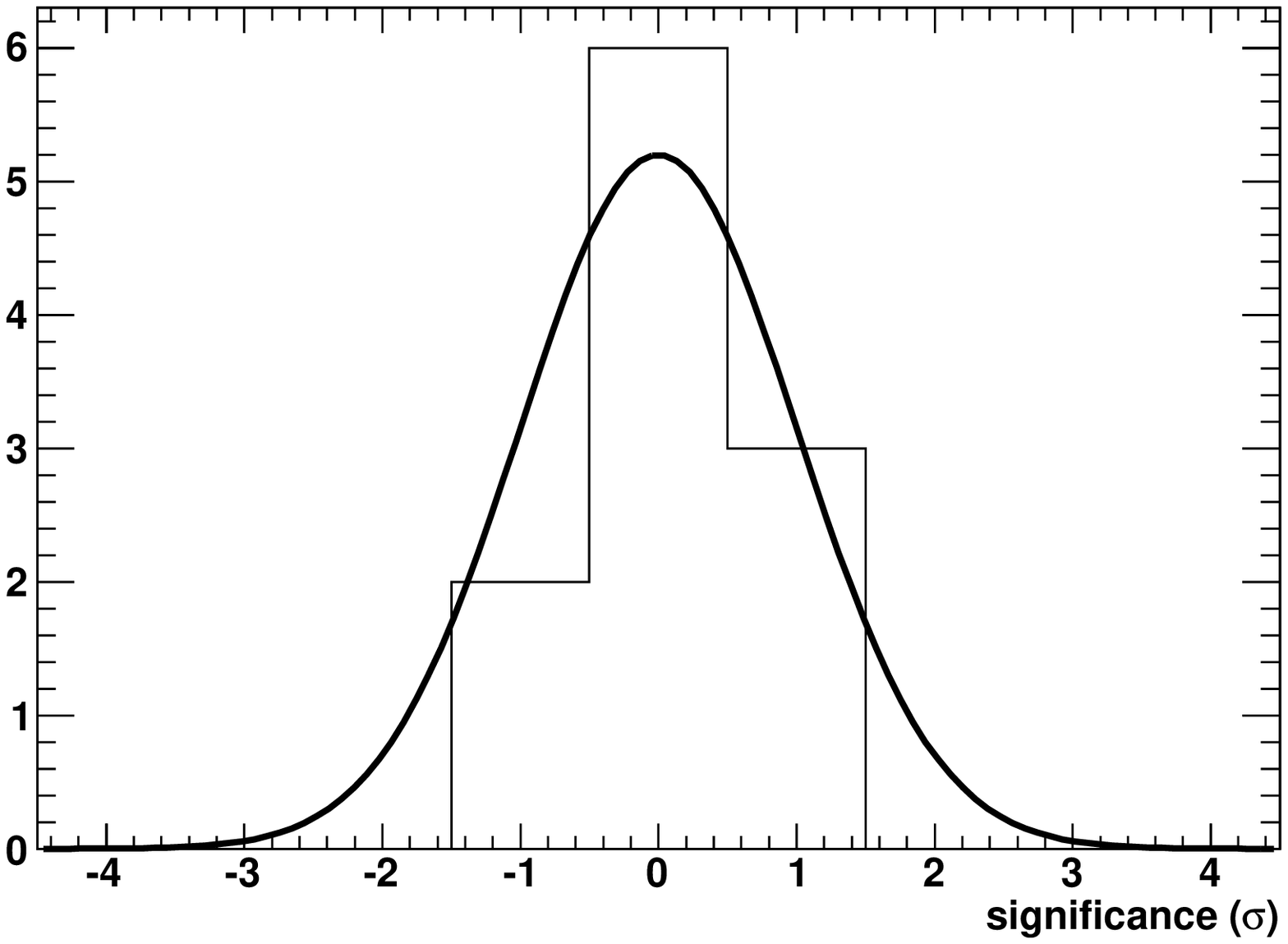}}
    }
  \end{center}
  \caption{Significance histograms obtained from an analysis of the GRBs in the sample over
    timescales for which VERITAS is maximally sensitive to a burst with a $t^{-1.5}$ power-law
    afterglow. Both standard and soft source analyses were performed. Included in the figures is the
    normalized Gaussian distribution of mean zero and variance one that the significance histograms
    should follow if no signal is present. The GRB significances are consistent with having been
    drawn from the aforementioned Gaussian distribution.}
  \label{fig:sigdistroOpttime}
\end{figure*}

The search for VHE gamma rays over timescales optimized for VERITAS sensitivity to a source with
$\frac{dN}{dE} \sim t^{-1.5}$ behavior was performed as described in the previous
section. Table \ref{tab:temporaltable} shows the results of this search. No emission associated with
any GRB in the sample of 16 presented in this paper is found. The distributions of
significances for both the soft and standard optimum time analyses are shown in Figure
\ref{fig:sigdistroOpttime}. For five of the bursts, the maximally sensitive duration of observation
is greater than the length of time spent observing the burst and these bursts are omitted from this
analysis. This occurred when the delay to the beginning of VERITAS observations was sufficiently
long.

\begin{table*}
  \begin{small}
    \caption{A Search for VHE Emission on Timescales Optimized on VERITAS Sensitivity to a Power Law
      Afterglow Decay $\sim t^{-1.5}$.}
    \label{tab:temporaltable}
    \begin{centering}
      \begin{tabular*}{\textwidth}{@{\extracolsep{\fill}}cc|ccccc|ccccc}
        \hline \hline
        \multicolumn{2}{c|}{} & \multicolumn{5}{c|}{Standard Source Analysis} & 
        \multicolumn{5}{c}{Soft Source Analysis} \\
        GRB & Duration(s) & $N_{\rm{on}}$ & $N_{\rm{off}}$ & $\sigma^{\alpha}$ & 
        $E_{\rm{th}}$ (GeV) & Upper Limit & 
        $N_{\rm{on}}$ & $N_{\rm{off}}$ & $\sigma^{\alpha}$&
        $E_{\rm{th}}$ (GeV) & Upper Limit \\
        \hline
        070223 & $2.7 \times 10^4$ & -- & -- & -- & -- & -- & -- & -- & -- & -- & -- \\
        070419A & $477$ & $2$ & $14$ & $0.8$ & $720$ & $4.0 \times 10^{-11}$ & $2$ & $42$ & $-0.9$ & $420$ & $4.6 \times 10^{-11}$ \\
        070521 & $1809$ & $3$ & $113$ & $-1.7$ & $170$ & $3.1 \times 10^{-12}$ & $23$ & $364$ & $-0.9$ & $110$ & $1.6 \times 10^{-11}$ \\
        070612B & $325$ & $3$ & $21$ & $0.9$ & $470$ & $3.8 \times 10^{-11}$ & $7$ & $58$ & $1.1$ & $270$ & $9.3 \times 10^{-11}$ \\
        071020 & $8509$ & -- & -- & -- & -- & -- & -- & -- & -- & -- & -- \\
        080129 & $2356$ & -- & -- & -- & -- & -- & -- & -- & -- & -- & -- \\
        080310 & $553$ & $3$ & $23$ & $-0.2$ & $480$ & $3.2 \times 10^{-11}$ & $13$ & $55$ & $1.4$ & $290$ & $7.9 \times 10^{-11}$ \\
        080330 & $252$ & $0$ & $15$ & N/A$^{\zeta}$ & $260$ & $2.4 \times 10^{-11}$ & $6$ & $43$ & $-0.2$ & $170$ & $1.4 \times 10^{-10}$ \\
        080409 & $1.1 \times 10^4$ & -- & -- & -- & -- & -- & -- & -- & -- & -- & -- \\
        080604 & $455$ & $2$ & $40$ & $-0.6$ & $200$ & $1.5 \times 10^{-11}$ & $9$ & $128$ & $-0.3$ & $140$ & $3.6 \times 10^{-11}$ \\
        080607 & $298$ & $4$ & $16$ & $1.1$ & $390$ & $9.2 \times 10^{-11}$ & $7$ & $46$ & $0.3$ & $250$ & $1.1 \times 10^{-10}$ \\
        081024A & $242$ & $1$ & $7$ & $-0.4$ & $270$ & $9.9 \times 10^{-11}$ & $4$ & $29$ & $0$ & $190$ & $1.9 \times 10^{-10}$ \\
        090102 & $8647$ & -- & -- & -- & -- & -- & -- & -- & -- & -- & -- \\
        090418A & $422$ & $3$ & $16$ & $0.4$ & $190$ & $3.1 \times 10^{-11}$ & $8$ & $46$ & $0.4$ & $120$ & $6.9 \times 10^{-11}$ \\
        090429B & $228$ & $2$ & $7$ & $0.8$ & $200$ & $9.9 \times 10^{-11}$ & $4$ & $27$ & $0.1$ & $140$ & $1.5 \times 10^{-10}$ \\
        090515 & $576$ & $4$ & $24$ & $0.3$ & $320$ & $2.7 \times 10^{-11}$ & $11$ & $72$ & $0.8$ & $210$ & $6.2 \times 10^{-11}$ \\
        \hline\hline
      \end{tabular*}
    \end{centering}
  \end{small}
  Upper limits defined as in Table 2.
  $^{\alpha}$Due to the low statistics, the calculation of the Gaussian significance by Equation 17 of
  \citealt{1983ApJ...272..317L} is not valid. The ratio of Poisson means, as discussed in
  \citealt{Cousins:2008p547} and \citealt{Zhang:1990p653} is employed instead, though it should be
  noted that the ratio of Poisson means method is quite conservative in situations with low
  statistics.  $^{\zeta}$In the case of zero ``on'' counts, the corresponding Gaussian significance is
  not defined.
\end{table*}

No significant excess of VHE gamma-ray events coincident with the large X-ray flare corresponding to
the interval $T_{\rm trig} + 475$ s to $T_{\rm trig} + 750$ s during the afterglow of GRB\,080310 (see Figure
\ref{fig:GRB080310XRTLC}) is found. After accounting for gamma-ray attenuation by the EBL, the soft
source analysis returns an integral upper limit of $9.8 \times 10^{-8}$ photon cm$^{-2}$ s$^{-1}$ above $310$ GeV. 
Though the flare was quite bright in the XRT band, increasing by $\sim 3$ orders of magnitude relative to the 
underlying afterglow, the burst was at a moderate redshift ($z=2.4$) so the VHE gamma-ray attenuation is 
presumably significant.

\section{Discussion}
The upper limits on VHE emission presented here provide strong constraints on the amount of energy
released during the early afterglow phase of GRBs. The limits themselves, however, are not
sufficient to reveal much without taking into context the effects of the EBL and the intrinsic
properties of each GRB. The nine bursts with measured redshifts have a mean and median redshift of
1.6 and 1.7, respectively. Assuming an EBL model \citep{2009MNRAS.399.1694G}, one may convert the
upper limits obtained from the VERITAS observations to limits on the intrinsic GRB flux as is done
in Table \ref{tab:redshiftTable}. The GRBs without measured redshifts are of less utility but, as a
first approximation, one may assume a redshift of $z = 2.5$, which is the approximate median of all
of GRBs with known redshift detected by {\it Swift} \citep{Gehrels:2009p4268}, to correct for the
gamma-ray attenuation from the EBL.

\begin{figure*}[htp]
\begin{center}
  \subfigure[]{\label{fig:edge-a}
    \scalebox{0.47}{
      \plotone{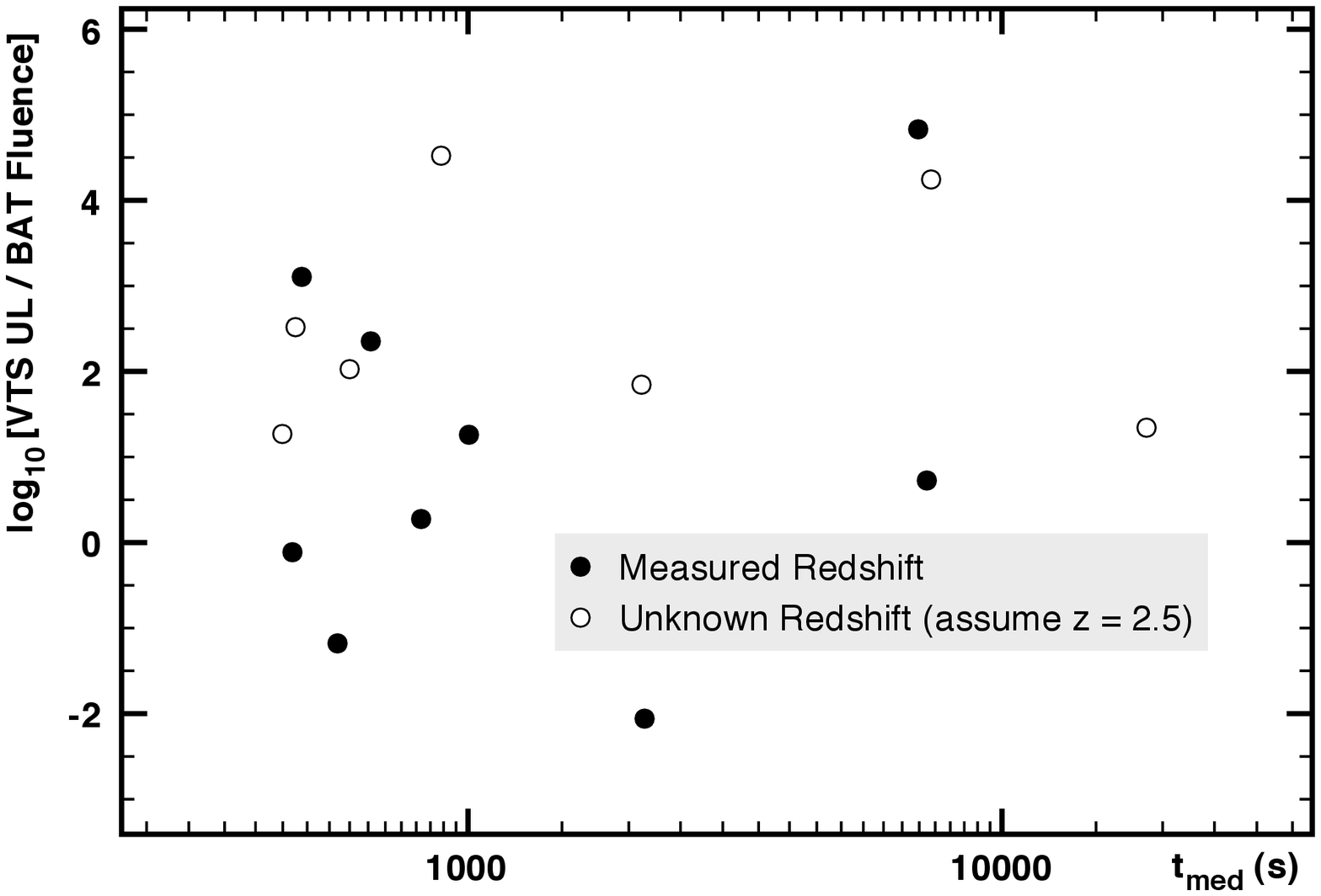}}
  }
  \subfigure[]{\label{fig:edge-b}
    \scalebox{0.47}{
      \plotone{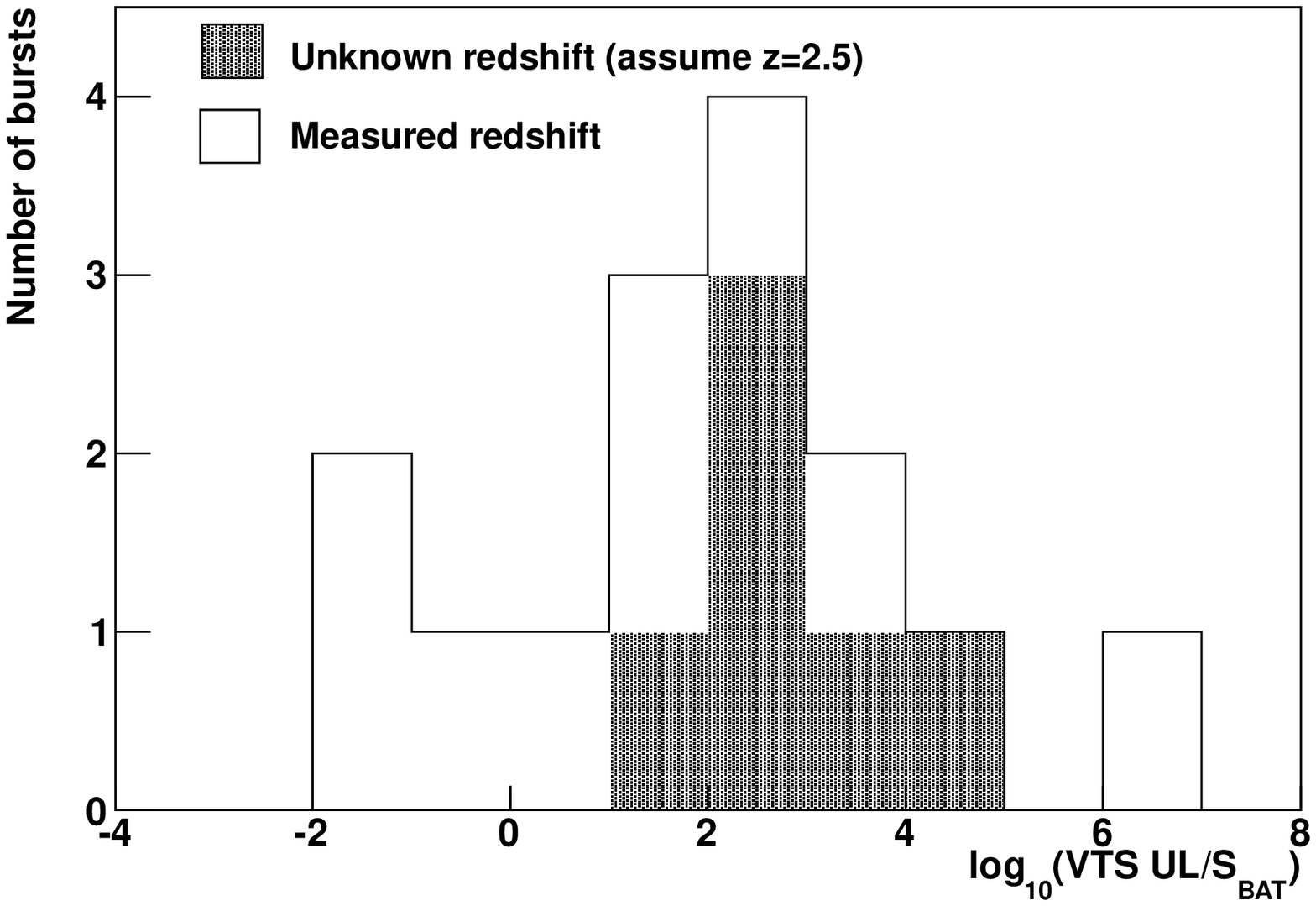}}
  }
  \end{center}
  \caption{(a) EBL-corrected VERITAS integral fluence upper limits above 200 GeV, divided by the
    fluence measured by the {\it Swift}-BAT in the 15 -- 350 keV energy band as a function of
    $t_{\rm{med}}$ as defined in the text. (b) A histogram of the ratio of the VERITAS integral
    fluence upper limit above 200 GeV, now integrated over the time period $t-T_{\rm trig}>300$ s, to
    the Swift-BAT fluence. One burst, GRB\,080409 is omitted in this plot as the VERITAS upper limit
    on the fluence of this burst is 11 orders of magnitude above the fluence measured by the
    BAT.}
  \label{fig:BATVHE}
\end{figure*}

After the VERITAS upper limits are corrected for EBL effects, we compare the VHE upper limits on the
fluence above 200 GeV with the fluence of the GRB as measured by the {\it Swift}-BAT in the $15$ --
$350$ keV energy range \citep{Butler:2007p4505,Butler:2010p4504} that is taken as a proxy for the
overall intensity of the burst. To account for the different delays and durations of the VERITAS
observations, we calculate $t_{\rm{med}}$, the time since the beginning of the VERITAS observations
of the GRB at which we expect to detect half of the photon signal, assuming a time profile of the
GRB afterglow of $\frac{dN}{dE} \propto t^{-1.5}$ that is motivated by the high-energy afterglows
observed by the {\it Fermi}-LAT. The ratio of VERITAS upper limit on the fluence above 200 GeV to
the BAT fluence versus $t_{\rm{med}}$, is plotted in the left panel of Figure \ref{fig:BATVHE} for
each burst. Since we assume a known time-dependence of the VHE afterglow, we may calculate this
ratio for any time period after the start of the GRB, which we take to be $t-T_{\rm trig}>300$ s. Then
for each GRB, we calculate the fractional upper limit on the VHE gamma-ray fluence over the
entire afterglow ($300 < t-T_{\rm trig} < \infty$) relative to the fluence measured by the BAT. A
histogram of this quantity is plotted in the right panel of Figure \ref{fig:BATVHE}. It should be
noted that if the bursts with unknown redshift are assumed to have the mean redshift of the GRBs in
our sample ($z=1.7$) as opposed to mean redshift detected by {\it Swift} ($z=2.5$), then the
distribution of bursts with unknown redshift moves to the left and more closely follows the
distribution of known-$z$ bursts.

These results show that for several bursts the VHE component of the GRB afterglow is less than the
energy released in the {\it Swift}-BAT band during the prompt phase of the burst. With observation
delays often on the order of a few hundred seconds, the VERITAS upper limits begin to restrict
theoretical models in which the afterglow from the forward external shock contains an SSC component
in addition to the synchrotron component \citep{2009ApJ...703...60X}.

VERITAS observations taken contemporaneously with X-ray flares during GRB afterglows are also of
interest. Over the time period of the flare observed during the afterglow of GRB\,080310, the
VERITAS upper limits constrain the integral of $F_{\nu}$ above $300$ GeV to be less than $9.4 \times
10^{-8}$ erg cm$^{-2}$ s$^{-1}$, which is a factor of $\sim 12$ above the peak flux observed by the {\it Swift}-XRT
in the 0.3 -- 10 keV band. 
In light of the fact that GRB\,080310 was at a redshift of nearly
$2.5$, it is clear that VHE observations of a strong X-ray flare from a low redshift GRB could
challenge some models in which SSC processes produce VHE emission simultaneously and with comparable 
intensity to the X-ray emission during the flare \citep{2008MNRAS.384.1483F} and add detail to our understanding of the processes occurring in GRB afterglows.

\section{Future Prospects}
The detection of VHE emission from GRBs in light of recent observations by the {\it Fermi} and {\it
  Swift} space telescopes remains a challenging, though not unreasonable, prospect. The number of
GRBs found by the LAT to emit GeV radiation is small, with a detection rate on the order of one
every few months. Combined with the $\sim$10\% -- 15\% duty cycle of an IACT array such as VERITAS,
the probability of simultaneous observation of such bursts is not high. On the other hand, $>$30 GeV
emission has been detected from both short \citep{2009Natur.462..331A} and long
\citep{2009ApJ...706L.138A} GRBs and, in the latter case, persists well after the prompt phase of
the burst. Furthermore, these observations indicate that the high-energy photon absorption due to
the EBL is not so severe \citep{Abdo:2010p2950} as to rule out ground-based VHE detections that in
turn could strongly constrain models of GRB physics \citep{Cenko2011}, as well as those of
the EBL.

\begin{figure*}[htp]
  \begin{center}
    \includegraphics[scale=0.85]{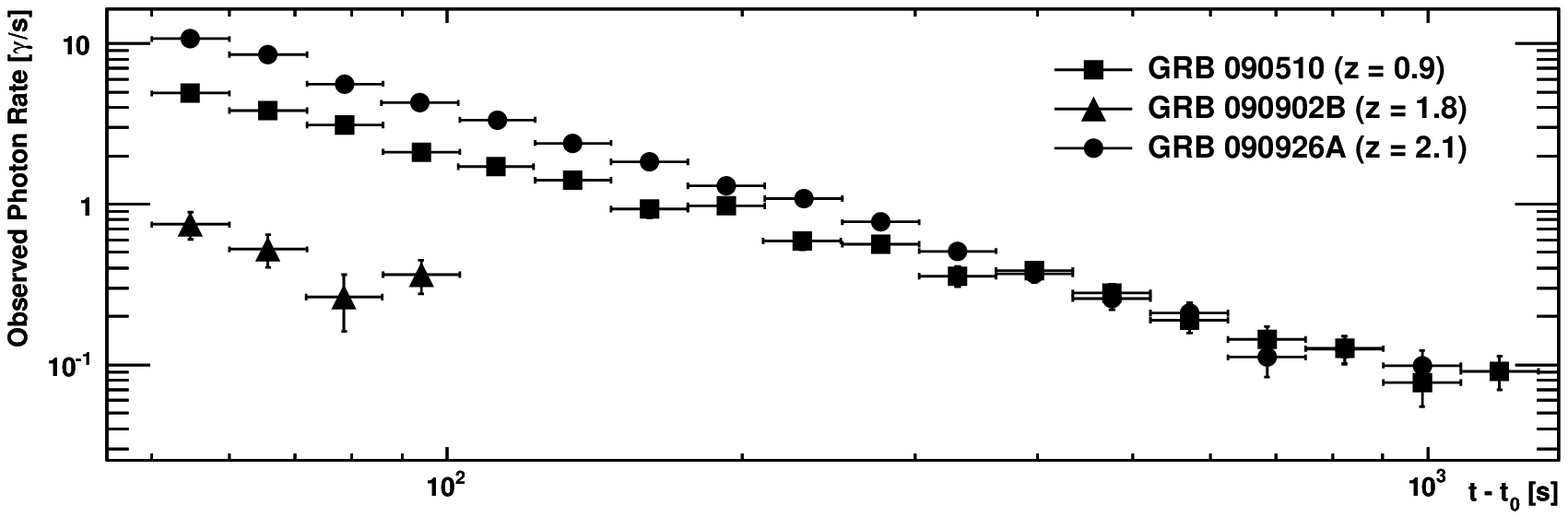}
  \end{center}
  \caption{Predicted VERITAS light curves for three of the four brightest {\it Fermi}-LAT GRBs. The
    fourth, GRB\,080916C had a redshift of nearly 4.4 and VHE emission is predicted to be too
    attenuated by the EBL to be detectable by VERITAS. The EBL model of
    \citet{2009MNRAS.399.1694G} is used to estimate the attenuation of the VHE $\gamma$-rays. The
    elevation of the burst with respect to VERITAS is chosen to be $70\degree$ and no intrinsic
    spectral cutoff of the high energy emission is assumed. Each point signifies a detection of at
    least three standard deviations ($\sigma$) in that time bin. }
  \label{fig:LATGRBLC}
\end{figure*}

Approximately one of every fifteen GRBs detected by the {\it Fermi}-GBM is detected by the LAT
(provided the burst also falls in the LAT FOV). Though they are rare, some luminous, LAT-detected
GRBs should be detectable by VERITAS. Taking the spectral and temporal characteristics of the high
energy emission from the four brightest {\it Fermi}-LAT bursts: GRB\,080916C
\citep{2009Sci...323.1688A}, GRB\,090510 \citep{DePasquale:2010p2521}, GRB\,090902B
\citep{2009ApJ...706L.138A}, and GRB\,090926A \citep{2011arXiv1101.2082F} we estimate the expected
flux of VHE photons in the energy range of VERITAS as a function of time. Figure \ref{fig:LATGRBLC}
shows the light curves of three of these four bursts that we predict to have been detectable by
VERITAS. GRB\,090510 and GRB\,090926A produce significant signal in the VERITAS band for roughly a
thousand seconds. GRB\,080916C had a redshift of $z>4$ and the VHE emission is extremely suppressed
through interaction with the EBL. It is observed that even for bursts with redshift between 1 and 2,
some exceptional GRBs may be quite bright in the VERITAS energy band. However, in the absence of
delayed activity (e.g. that associated with X-ray flares) the power law temporal decay of the
late-time, high-energy emission necessitates relatively rapid follow-up observations. VERITAS has
made several GRB follow-up observations with delays of less than 100 s and has a median
response time of 328 s\footnote{These numbers are based on all GRBs observed by VERITAS from
  2007 January through 2009 June, including {\it Fermi}-GBM triggered observations that are not
  included in this paper.}  and therefore may be capable of detecting the same high-energy component
that the {\it Fermi}-LAT detects, provided it extends to $>100$ GeV energies.

VERITAS continues to take follow-up observations of GRBs. In the summer of 2009 one of the
telescopes in the VERITAS array was moved to a new position that resulted in an improvement in
sensitivity of $\sim 30 \%$.  By fall 2012, an upgrade of the telescope-level trigger system and the
replacement of existing PMTs with a more sensitive PMT will significantly increase the low energy
response of the instrument. This is particularly important for GRB observations as the EBL
significantly attenuates the high-energy component of sources with appreciable
redshifts. Additionally, work is ongoing to improve the sensitivity of the array with respect to low
elevation targets, which make up the majority of GRB observations. Response times for immediately
observable bursts have been gradually decreasing and efforts are underway to increase the slewing
speed of the telescope motors to reduce these times further. Such efforts are increasing the GRB
science capability of VERITAS and will lead to a more thorough characterization of the highest
energy emission from GRBs.

\section{Conclusions}
The VERITAS telescope array was used to make follow-up observations of 29 satellite-detected GRBs
during the period of 2007 January through 2009 June. Due to the incorporation of real-time alerts
from the GCN into the VERITAS pointing and control software, relatively small observation delays
(down to 91 s) were achieved. After quality selection cuts, data from 16 of the 29 bursts were
analyzed and those results are presented here. No significant excess of VHE gamma rays from any of
the bursts is detected and the 99\% confidence level upper limits on the photon flux are
derived. Assuming a $t^{-1.5}$ temporal decay of the VHE afterglow, limits on the VHE afterglow
fluence relative to the prompt fluence detected by the {\it Swift}-BAT are calculated. For more than
half of the GRBs with known redshift in our sample, the VHE afterglow fluence is constrained to be
less than the prompt, low-energy gamma-ray fluence.

In the context of recent GRB observations by {\it Fermi}-LAT, prospects for detection of VHE
emission by VERITAS are good, assuming the most straightforward extrapolation of the spectral and
temporal characteristics of the high-energy emission. Contemporaneous early-afterglow observations
of a GRB by {\it Fermi}-LAT and an IACT array would provide valuable insights into understanding the
physical processes at work in the GRB environment as well as constrain the properties of the
EBL.

\acknowledgments This research is supported by grants from the US Department
of Energy, the US National Science Foundation, and the Smithsonian Institution, by NSERC in Canada,
by Science Foundation Ireland (SFI 10/RFP/AST2748), and by STFC in the UK. We acknowledge the
excellent work of the technical support staff at the FLWO and the collaborating institutions in the
construction and operation of the instrument. We acknowledge the support provided by NASA {\it
  Swift} Guest Investigator (GI) grant NNX09AR06G. T. Aune gratefully acknowledges the support
provided by a NASA Graduate Student Researchers Program fellowship. This work made use of data
supplied by the UK Swift Science Data Centre at the University of Leicester.

{\it Facilities:} \facility{VERITAS}.

\end{document}